# Cost overruns in Large-Scale Transportation Infrastructure Projects: Explanations and Their Theoretical Embeddedness

By

Chantal C. Cantarelli, Bent Flybjerg, Eric J. E. Molin, and Bert van Wee

**Abstract**

Managing large-scale transportation infrastructure projects is difficult due to frequent misinformation about the costs which results in large cost overruns that often threaten the overall project viability. This paper investigates the explanations for cost overruns that are given in the literature. Overall, four categories of explanations can be distinguished: technical, economic, psychological, and political. Political explanations have been seen to be the most dominant explanations for cost overruns. Agency theory is considered the most interesting for political explanations and an eclectic theory is also considered possible. Non-political explanations are diverse in character, therefore a range of different theories (including rational choice theory and prospect theory), depending on the kind of explanation is considered more appropriate than one all-embracing theory.

# Introduction

Investments in infrastructure are a considerable burden on a country's gross domestic product (GDP). For example, in 2005 the Dutch government invested about 8 billion euros (CBS, 2005 in KIM, 2007) in infrastructure, amounting to 1.55% of GDP. This is of even greater concern if the inefficient allocation of financial resources as the result of decisions based on misinformation are recognised (Flyvbjerg, 2005b, De Bruijn and Leijten, 2007). Cost estimates are often inaccurate and consequently the ranking of projects based on project viability is also inaccurate. Inevitably, this means there is a danger that eventually inferior projects are implemented, that resources are used which could have been assigned more appropriately, and that projects that are unable to recover their costs are implemented. Inaccurate estimates make it particularly difficult to manage large projects and often lead to cost overruns, which further increases the burden on the country's GDP. The problem can be summarised as follows: managing large-scale transportation infrastructure projects is difficult due to frequent misinformation about the costs which results in large cost overruns that often threaten overall project viability. Various studies have addressed the issue of cost overruns in transportation projects (van Wee, 2007). Some studies, including a large database of projects, reach the following conclusions. The Government Accountability Office, for example, found that 77% of highway projects in the USA experienced cost escalation (in Kaliba et al., 2008). Merewitz (1973) suggests that the average overrun of infrastructure projects is a little over 50 percent (Merewitz, 1973). A review by Morris and Hough (1987), which covered about 3500 projects, revealed that overruns are the norm, and generally range between 40 and 200 per cent (Reichelt and Lyneis, 1999). Furthermore, a study by Flyvbjerg et al. (2003a) indicates that in 86 percent of the projects cost overruns appear to overrun by an average of 28 percent. The problem is recognised in the literature but the causes and explanations are still ambiguous. To the authors' knowledge, a systematic investigation into the different explanations for cost overruns has not yet been conducted. Moreover, insight into the theories underlying these explanations has been the subject of only a few studies. A sound theoretical basis is particularly important because it substantiates the explanation and provides opportunities to define the appropriate cures.

This paper provides an overview of explanations and their theoretical embeddedness in order to gain a better understanding of the phenomenon of cost overruns

The paper is structured as follows. The second section describes the research methodology, and this is followed in section 3 by a description of the causes and explanations for cost overruns for each source. The explanations are categorised and further examined in section 4. Section 5 elaborates on the theoretical embeddedness of the explanations. Finally, section 6 presents the main conclusions, addresses the research questions and presents a number of recommendations.

**Methodology**

In line with the conventional methodology, the inaccuracy of cost estimates is measured as the size of cost overruns. *Cost overrun* is measured as actual out-turn costs minus estimated costs as a percentage of estimated costs. Actual costs are defined as real, accounted construction costs determined at the time of project completion. Estimated costs are defined as budgeted or forecasted construction costs determined at the time of the decision to build. Cost estimates become more accurate during the project process. However, what is relevant here is the estimate known by the decision maker, i.e. the estimate based upon which the decision maker decides whether or not to implement the project. A particular moment in time is often taken to represent the moment at which the decision to implement the project was made ('formal decision to build'). Cost overruns are generally calculated according to the costs estimated at this 'formal decision to build' (these are the costs at the initial funding level). However, the decision-making process involves several moments at which decisions are made; therefore, references to the formal decision to build do not always provide an accurate picture of cost overruns. In some cases, parties have committed themselves at an earlier decision-making moment, known as the 'real decision to build'. This situation is referred to as *lock-in* at the decision-making level. Lock-in influences the magnitude of cost overruns, because the estimated costs at the real decision to build are usually lower than those at later stages of the decision-making process (Cantarelli et al. 2009). This paper concentrates on *explanations* rather than on *causes*. In line with the definition in Flyvbjerg et al. (2004), what we mean by 'cause' is 'to result in'; the cause is not the explanation of the result. Causes refer to the variables or factors that influence the cost overruns, such as the implementation period or the size of the project. Explanations are more general and might comprise several causes.

We define transportation infrastructure projects as follows: 'Transport infrastructures include roads, rail lines, channels, (extensions to) airports and harbours, bridges and tunnels. Of these projects it is the 'hardware' that is considered, and the "software",i.e. projects relating to deregulations, liberalization, privatization, and so forth is excluded'. The literature did not provide one minimum cost level that is generally applied to mark a large-scale project. A large-scale project is defined in this paper by a minimum cost level of 500 million euros.

A literature study of explanations and theories that are used to support the explanations was carried out. In line with Morris (1990), who concludes in his research that in understanding the planning failures, one has to look for a general explanation, the review methodology does not set out any restrictions in the search for literature on cost overruns of transportation infrastructure projects. It attempts to give an overview of studies that is as complete as possible. Studies addressing project performance in general are considered (broad focus) as well as studies focusing specifically on cost overruns (narrow focus). Most studies are empirical studies; studies that largely use data from observation or experience, i.e. empirical studies give insight into the extent of cost overruns based on data from real projects. Table 1 presents the different studies.

**Table 1. Overview sources of literature**

|  | Various categories of projects including transport projects | Transport |
|---|---|---|
| Narrow focus | Wachs (1987, 1989) | Knudsen (1976) |
|  | Morris (1990) | Fouracre et al. (1990) |
|  | Arvan and Leite (1990) | Pickrell (1992) |
|  | Kahneman (1993, 2003) | Auditor General of Sweden (1994) |
|  |  | Mansfield et al. (1994) |
|  |  | Skamris and Flyvbjerg (1997) |
|  |  | Nijkamp and Ubbels (1999) |
|  |  | Trujillo (2002) |
|  |  | Odeck (2004) |
|  |  | Lee (2008) |
|  |  | Kaliba et al. (2008) |
| Broad focus | Hall (1980) | Szyliowics (1995) |
|  | Altshuler and Luberoff (2003) | Bruzelius et al (2003) |
|  |  | Flyvbjerg et al. (2003) |
|  |  | Mackie and Preston (1998) |

## Causes and explanations for cost overruns

### Studies with a narrow focus

*Morris (1990)* conducted one of the first empirical studies with a narrow focus on cost overruns in large projects. He argues that delays in project implementation and cost overruns have become a regular feature of *public* sector projects. The average cost overrun found in this study is 82%. As far as possible causes are concerned, Morris (1990) concludes that about 20 - 25% can be attributed to price increases, and the remaining 70-75% has to be explained in terms of real factors, such as delays in implementation. He gives the following main factors as the causes of delays and cost overruns: poor project design and implementation, inadequate funding of projects, bureaucratic indecision, and a lack of coordination between enterprises.

The study by *Arvan and Leite (1990)* focuses on large-scale government sponsored procurement. They provide an explanation of cost overruns by assuming that the sponsor cannot pre-commit to the compensation paid to the contractor when the contractor has some private cost information.

*Wachs (1987, 1989)* reviews several forecasting models in the field of transportation. He finds that forecasts are often inaccurate, underestimating costs and overestimating traffic demand. He proposes two possible explanations for these optimistic forecasts. Firstly, 'forecasting is inherently exact and the observed errors result from imperfect techniques'. Secondly, 'travel and cost forecasting is deliberately slanted to produce figures which constitute technical justification for public works programs favoured on the basis of political rather than economic or technical criteria'. Because the forecasting errors are always in the same direction - always an overestimation of traffic demand and an underestimation of costs - the first explanation seems, according to Wachs, to be less valid. In line with Ascher's argumentation (1987) he concludes that 'the competitive, politically charged environment of transportation forecasting has resulted in the continuous adjustment of assumptions until they produce forecasts which support politically attractive outcomes'. He identifies three main sources of error in forecasting costs: changes of scope, assumed rates of inflation that are lower than actual rates of inflation, and delay. He concludes that about 40-90% of the total cost overrun can be explained by these factors, but a substantial part remains unexplained. Other causes can be found in the funding system commonly found in rail transit projects. There is an incentive with this kind of funding system to select the most optimistic assumptions in the development of cost estimates for projects.

A frequently cited piece of research concerned with forecasting in decision-making is by the Nobel prize winner *Kahneman*. Kahneman and Lovallo (1993) and Lovallo and Kahneman (2003) identify two main biases in forecasting and risk taking. The first bias concerns optimism bias, the systematic tendency to be overly optimistic about the outcome. The second bias concerns risk aversion, the overly cautious attitudes towards risk.

Lastly, a more recent study by Lee (2008) examined cost overruns in Korean social overhead capital projects. Based on 161 completed projects he concluded that the causes of cost overruns can be grouped into several major categories: changes in scope, delays during construction, unreasonable estimation and adjustment of project costs, and no practical use of the earned value management system.

Various studies addressed cost overruns for *transportation projects* specifically. For example, *Pickrell (1992)* investigated the cost overruns and benefit shortfalls of 8 rail transit projects in the US. In his study, Pickrell (1992) starts from the premise that forecasters overestimate rail transit ridership and underestimate rail construction costs and operating expenses. To understand these inaccurate forecasts, he points, on the one hand, to optimism among local officials and to inadequate planning processes on the other. He argues that the causes of underestimated costs lie in the structure of programmes and the existence of dedicated funding sources that provide few incentives for local officials to seek accurate information for evaluating alternatives. *Fouracre et al. (1990)* investigated cost overruns for 21 metro projects worldwide. Nearly all the metro systems incurred costs higher than expected. These overruns were attributed to 'a range of factors, including the additional costs of unforeseen service and utility diversions and other civil works problems, which could not be offset by contingency allowances; changes in specifications; currency devaluation and rises in interest charges'. According to the authors, most of the cost estimates were optimistic because there was little appreciation of the difficulties of the work. In addition, authorities lacked the management skills to mitigate errors in project planning and to keep effective control of costs.

The *Auditor General of Sweden (1994)* is another study with a narrow focus on cost overruns involving transport projects. It covered 15 road and rail projects. The average capital cost overrun for the eight road projects was 86%, ranging between 2 and 182%, and for the seven rail projects this was 17%, ranging from minus 14% to plus 74%. The authors conclude that there is still a considerable element that cannot be explained by technical causes.

The study by *Mansfield et al. (1994)* considered the causes of cost overrun in Nigerian construction projects specifically (highway projects). They concluded that the major variables that can lead to excessive project overruns are the financing of and payment for completed works, poor contract management, shortages of materials, price fluctuations, and inaccurate estimates leading to delays. Other factors which can be identified as usually being responsible for project delays and excessive costs are excessive bureaucratic checking and approval procedures, unclear definitions of contract terms by the client and insufficient geotechnical investigations at the feasibility stage.

The research by *Skamris and Flyvbjerg (1997)* covers seven tunnel and bridge projects. They found an average construction cost overrun for the five completed projects of 14%, ranging from -10% to 33%.

The Dutch study by *Nijkamp and Ubbels (1999)* also concentrates specifically on the cost overruns of transport projects. In contrast to the findings of most studies, they conclude that in cost estimates generally tend to be rather reliable. In most projects, cost overruns were common but the extent of cost underestimation varied between 0 and 20%. They identify three common causes of cost underestimation in projects: price rises, incompleteness of estimations and adjustments to the projects. They do not consider the strategic behaviour of the actors involved to have a major impact on cost overruns. They tend to argue that change in social opinion and intervention by interest groups, the availability of new technologies, the state of the economy, and the tendering method all lead to adjustments in the project which cause cost overruns.

A more recent study on cost overruns by *Odeck (2004)* uses statistical analysis to derive the average cost overruns and to identify the factors that influence cost overruns. The average cost overrun found in this study is rather small at around 7.9%. A striking feature is the large standard deviation – 29.2% – indicating a large spread around this average among the individual projects. Surprisingly, the cost overrun percentage seems to be higher for smaller projects compared with larger ones. (However, the number of large projects is small compared with the number of smaller projects.) Regarding the factors that influence cost overruns, it was concluded that completion time and the geographical region influence cost overruns, whereas project type and workforce do not have an impact. Odeck (2004) argues that larger projects are most probably under much better management compared with smaller ones and this is the reason why overruns are less predominant among larger projects. As a possible explanation for the tendency that cost overruns are higher the shorter the completion time, he argues that the shorter the length of time the construction is expected to take, the more difficult it is to predict costs. This would imply that uncertainties diminish with time.

Kaliba et al. (2008) carried out a study into cost escalation and schedule delays in road construction projects in Zambia. The main causes of cost escalation were: bad or inclement weather due to heavy rain and flooding, scope changes, environmental protection and mitigation costs, schedule delay, strikes, technical challenges, inflation and local government pressure. Factors that lead to cost escalation are said to include: the size of the project; project scope enlargement; inflation; length of time to complete the project; incompleteness of preliminary engineering and quantity surveys; engineering uncertainties; exogenous delays; complex administrative structures; and inexperienced administrative personnel (Merewitz, 1973). Cost escalation is further compounded by factors such as project location, project conditions, environmental mitigation costs, suspension of work, strikes, poor site coordination, expiry of bid, local government pressure, political discontinuity and transportation problems (Hall, 1980; NAP, 2003; Schexnayder, 2003).

**Studies with a broad focus**

*Hall (1980)* conducted one of the first empirical studies with a broad focus on inadequate planning of large infrastructure projects incorporating cost overruns. The research starts with the notion that many of the planning disasters seem to have been initiated on the basis of forecasts that were later found to be inadequate and misleading. Searching for a better understanding of the failures in planning, Hall (1980) considers planning uncertainty to be an important element and makes a distinction between three categories of uncertainty. They are: uncertainty in the planning environment, uncertainty in related decision areas and uncertainty about value judgments (see: Hall, 1980, for an elaboration on these types of uncertainty). He further considers whether the difference between public and private goods has any effect on the planning failures. According to Hall (1980), the main problem is the way in which societies plan the output of the *public* good (goods and services which the public is willing to pay for but which the private sector is not motivated to provide (Hall, 1980)). Public goods are characterised by non-exclusiveness and non-control over exclusion (Snidal, 1979). Suppliers of the public good do not have the opportunity not to provide the good (non-exclusiveness). This difference between public and private goods is particularly important in the research on cost overruns.

Mackie and Preston (1998) present twenty-one sources of error and bias in the appraisal of transport projects. They mainly relate to measurement error and appraisal optimism. They conclude that appraisal optimism is the greatest danger in transport investment analysis. 'Appraisal optimism happens because the information contained in the appraisal tends to be owned by scheme promoters who have obvious incentives to bias the appraisal - deliberately or unwittingly'.

Another study that incorporates a wider scope is the research of *Bruzelius et al. (2002)* who find that differences between forecasts and actual costs, revenues and viability could not be explained by the difficulty of forecasting itself. These differences can only be explained by the strategic behaviour of project proponents who succeed in biasing forecasts in such a way that it leads to the decision to continue with the project instead of to change plans. Three issues are mentioned in this respect: the lack of a long-term commitment to the project, rent-seeking behaviour for special interest groups, and the tendency to underestimate in tenders in order to get proposals accepted.

Research by *Altshuler and Luberoff (2003)* focuses on the new politics of infrastructure development and distinguishes four political eras. One of the main important conclusions of the research relevant here is the following notion: 'consistent underestimation is an example of the tragedy of the commons. It corrodes the public confidence in government overall, and especially in proposals with long time frames, even as it helps advance specific projects'.

Finally, one of the leading pieces of research in the field of cost overruns in large transportation infrastructure projects is by *Flyvbjerg et al. (2003a)*. They examined 258 projects worldwide, and their research identifies cost overruns for several projects. They find that cost overruns are the greatest for rail projects, with an average cost overrun of 45%, followed by fixed links (average cost overruns of 34%) and road projects (average cost overrun of 20%). Explanations for cost overruns are sought through statistical analysis and theoretical considerations. Four categories of explanations were distinguished (see for example Flyvbjerg et al. 2002a, Flyvbjerg 2005, Flyvbjerg et al. 2003a). First, technical explanations are indicated, which are forecasting errors in technical terms, including inadequate data and lack of experience. Second, there are economic explanations that depict the cost underestimation as deliberate and economically rational. Third, psychological explanations for cost overruns, including the concepts of planning fallacy and optimism bias, are provided. Fourth, political explanations might also explain cost overruns. Strategic misrepresentation is an important concept within political explanation.

To obtain a better overview of the type of causes and explanations, section 4 will categorise these causes and explanations.

## Categorising causes and explanations

Table 2 presents the causes and explanations found in the studies considered based on the categorisation provided by Flyvbjerg et al. (2003a).

**Table 2. Causes and explanations**

| Explanation | Causes | Study |
|---|---|---|
| Technical | Forecasting errors including price rises, poor project design, and incompleteness of estimations | Morris, Nijkamp and Ubbels, Lee, Fouracre, Mansfield et al., Kaliba et al., Mackie and Preston |
| | Scope changes | Nijkamp, Wachs, Lee, Fouracre et al., Kaliba |
| | Uncertainty | |
| | Inappropriate organisational structure | Hall, Kaliba et al. |
| | Inadequate decision-making process | Hall, Mansfield et al., Kaliba et al. |
| | Inadequate planning process | Bruzelius et al. |
| | | Pickrell |
| Economical | Deliberate underestimation due to: | |
| | - lack of incentives, | Pickrell, Wachs |
| | - lack of resources, | Odeck, Mansfield et al. |
| | - inefficient use of resources | Hall |
| | - dedicated funding process | Pickrell, Morris, Wachs, Bruzelius et al. |
| | - poor financing / contract management | Mansfield et al. |
| | - strategic behaviour | Hall, Bruzelius et al. Arvan and Leite |
| Psychological | Optimism bias among local officials | Pickrell, Kahneman and Lovallo, Fouracre et al., Mackie and Preston |
| | Cognitive bias of people | Kahneman and Lovallo |
| | Cautious attitudes towards risk | Kahneman and Lovallo |
| Political | Deliberate cost underestimation | Nijkamp, Bruzelius et al. |
| | Manipulation of forecasts | Wachs, Auditor General of Sweden |
| | Private information | Arvan and Leite |

*Technical explanations* are commonly found in the literature on cost overruns. Price rises, poor project design and implementation, and incomplete estimations are all seen as the causes of cost overruns. Price rises are difficult to predict in the future, poor project design and implementation could be the result of a lack of experience, and incomplete estimates are an indication of inadequate data. These are considered variables that influence cost overruns, rather than explaining cost overruns themselves. Together with other causes, the cause is part of a technical explanation. Scope changes, uncertainty, inappropriate organisational structure, inadequate decision-making processes, and inadequate planning processes are all considered technical explanations for cost overruns on their own. They mainly relate to difficulties predicting the future and are considered 'honest' errors. Scope changes indicate changes in the design that were not predicted beforehand. These changes involve additional costs. The inappropriate organisational structure, the inadequate decision-making process, and the inadequate planning process all indicate inefficiency resulting in costs higher than expected. What we are looking at here are an inability to adapt sufficiently well to changing circumstances, accountability and control, and planning.

The lack of incentives and resources, the dedicated funding process, and the inefficient planning of public outputs are considered (economic) causes because although they influence the extent of cost overrun, they cannot provide an explanation in themselves. Forecasters often lack an incentive to provide accurate estimates and accordingly underestimate forecasts because it is in their own interest to do so. Due to a lack of resources, decision-makers have to choose between projects and this leads to competition. Consequently, project promoters deliberately underestimate costs in order to make projects look more attractive and thereby increase the chance of being selected. The inefficient use of resources can also result in cost overrun. Inferior projects are implemented and resources are spent that cannot be recovered. Lastly, the dedicated funding process results in cost overruns. Costs of projects are deliberately underestimated to increase the chance of receiving part of the funding. Strategic behaviour is an economic explanation for cost overruns on its own. Underestimating costs increases the chance of getting the project started.

*Psychological explanations* are based on the concepts of planning fallacy and optimism bias. They involve peoples' cognitive bias and their cautious attitudes towards risks When taking decisions. In taking decisions with risky prospects, people tend to be risk averse, have near-proportional risk attitudes (people are proportionally risk averse) and frame their decision problems narrowly (people consider decision problems one at a time, often isolating the current problem from other choices that may be pending, as well as from future opportunities to make similar decisions (Kahneman and Lovallo, 1993)). The cognitive bias leads to optimistic forecasts resulting in cost overruns. And due to the cautious attitude towards risks, people frame an outcome that maximises utility. A higher utility is obtained when the project is selected for implementation. The chance of being selected is increased when the estimated costs are low, consequently leading to underestimation.

*Political explanations* are generally agreed upon in the literature as the main explanation for cost overruns. Other explanations (sub-explanations) that fall within this overall category are deliberate cost underestimation and forecast manipulation. Costs are deliberately underestimated in order to increase the chances of project acceptance. Wachs (1989) argues that cost forecasts are manipulated because behaviour is determined on considerations of advocacy rather than objectivity. The literature furthermore describes different causes of cost overruns by strategic misrepresentation, including: learning, a lack of coordination, a lack of long-term commitment, a lack of discipline, organisational and political pressure, and asymmetric information. Learning involves the awareness among managers and decision-makers that in order for projects to be selected for implementation, forecasts of outcomes have to be highly favourable. Consequently, they behave strategically and misrepresent forecasts. The lack of coordination, the lack of long-term commitment and the lack of discipline make strategic behaviour possible because of the lack of consequences that is related to this kind of behaviour. Organisational and political pressures cause strategic misrepresentation because forecasts are adjusted to derive the most politically or organisationally attractive outcomes. Lastly, asymmetric information is an important cause of deliberate underestimation or strategic misrepresentation. Decision-makers have little information and are dependent on the information obtained from forecasts. This gives forecasters the opportunity to misrepresent information.

It is recognised within this categorisation of explanations that the difference between economic and political explanations is rather small. Both types of explanation use utility as a basis to understand behaviour. However, the starting point differs. Whereas economic explanations reason from the lack of incentives and resources and consider this the starting point to strive for utility maximisation, political explanations construe this in terms of interests and power (Flyvbjerg, 1998).

## Plausibility of explanations

The plausibility of an explanation is partly based on its theoretical embeddedness. When there are models, assumptions, premises or concepts behind the explanation, the likelihood of understanding the phenomenon of cost overruns increases.

Table 3. Theories in explanations

| Explanation | Theory | Study | Type of study |
|---|---|---|---|
| Technical | Forecasting | Kahneman and Lovallo, Wachs Flyvbjerg et al. | Narrow & various, Broad & transport |
| | Planning | Pickrell, Altshuler and Luberoff, Hall | Narrow & transport, Broad & various |
| | Decision-making | Bruzelius et al. | Broad & transport |
| Economical | Neoclassical economics | Pickrell, Odeck, Wachs | Narrow & transport, Narrow & various |
| | Rational choice | Hall, Flybjerg et al. | Broad & various, Broad & transport |
| Psychological | Planning fallacy & optimism bias | Kahneman and Lovallo, Pickrell, Flyvbjerg et al., Fouracre et al., Mackie and Preston | Narrow & various, Narrow & transportation, Broad & transport, Narrow &Transport |
| | Prospect | Kahneman and Lovallo, Flyvbjerg et al. | Narrow & various, Broad & transport |
| | Rational choice | Kahneman and Lovallo | Narrow & various |
| Political | Machiavellianism | Flyvbjerg et al., Bruzelius et al., Hall, Wachs, Morris, Pickrell , Nijkamp and Ubbels, Odeck | Broad & transport, Broad & various, Narrow & various, Narrow & transport |
| | Agency | Wachs, Flyvbjerg et al., Arvan and Leite | Narrow & various, Broad & transport, Narrow & various |
| | Ethical | Wachs, Flyvbjerg et al., Auditor General of Sweden | Narrow & various, Broad & transport, Narrow and transport |

Table 3 shows that a large variety of theories is used to support explanations. Theories are evenly distributed among studies.

**Technical explanations**

Three theories were used to support technical explanations: forecasting theory, planning theory and decision-making theory. Forecasting theory examines estimations in uncertain future situations. It studies the understanding of the forecasting process at large and aims to clarify how and why the various successes and failures come about (Armstrong, 2001). Failures in estimates may arise as a result of the cognitive mind in the forecasting process. Forecasting models were used to gain a better understanding of the problems with errors in forecasting techniques or inappropriate forecasting approaches that lead to poor cost estimates. Planning theory examines how projects and policy are established (Faludi, 1973). Planning concepts were used to refer to the inappropriate planning process of projects and the poor design and implementation as a main explanation for cost overruns. Lastly, *decision-making theory* considers government and politics as a series of decisions taken by people and institutions that make rational decisions in the light of their interests and the circumstances under which they operate (Dunleavy, 1991). This is mainly seen when it is referred to inappropriate institutional arrangements as a reason for cost overruns. The three theories are rather different and can be useful to address different parts of the explanation.

**Economic explanations**

Economic explanations were mainly founded on neoclassical economics and rational choice theory. Neoclassical economics is a framework for understanding the allocation of scarce resources among alternative ends. It sees that incentives and costs play an important role in shaping decision making. These notions of incentives in decision making are used in relation to cost overruns as follows: 'The dedicated funding causes little incentive to produce accurate figures because accurate figures decrease the chance of receiving part of the funding' (Pickrell, 1992). The premises of neoclassical economics are also used to find an explanation for the tendency to deliberately misrepresent information. This is explained by the lack of incentives for the planners in their role as 'advocates'. Rational choice theory aims to understand social and economic behaviour. It assumes that the actions of individuals are fundamentally rational and people calculate the costs and benefits of an action, recognising their preference functions and constraints facing them before taking a decision (Arrow, 1987;

Coleman, 1992). The theory is used to underlie the explanation that it is economically rational to underestimate costs because it will increase the likelihood of revenue and profit.

Rational choice theory is considered to have considerable potential in explaining cost overruns, not only for economic explanations but also for psychological and political explanations. For political explanations, it has important implications for the relation between the agent and the principal. The theory assumes that individuals choose the best action according to stable preference functions and the constraints facing them. When making a decision, the agent searches for the best action according to his preferences, taking the interests of the principal into account. This might lead to conflicts surrounding the cost estimates.

**Psychological explanations**

Psychological explanations are addressed by a small number of studies and are based on the concepts of planning fallacy and optimism bias, prospect theory and rational choice theory. *Planning fallacy* is used as follows: 'it is the tendency to underestimate time, costs and risks of future actions and at the same time overestimate the benefits of the same actions'. Cognitive biases of forecasters such as scenario thinking, anchoring estimations and extrapolation of current trends result in *optimism bias*, the systematic tendency to be overly optimistic. *Prospect theory* (which is part of psychological theory) is used to explain that the optimistic forecasts are a result of decision-making involving uncertainties and risk. The explanation of cost overruns based on risks can also be founded by *rational choice theory* which assumes that in their consideration people take risk into account in their goal of utility maximisation.

The concept of planning fallacy and optimism bias are closely related, but because the link with cost overruns is stronger for optimism bias, the preference is given to this notion to support psychological explanations. Prospect theory is preferred even more so because it provides a more comprehensive model for psychological explanations incorporating uncertainty and risks in addition to optimistic forecasts. Lastly, rational choice theory is considered a very useful basis for understanding cost overruns because it addresses economic, political and psychological elements of the phenomenon.

**Political explanations**

Three theories underlie political explanations: the concept of Machiavellianism, agency theory, and ethical theory. Strategic misrepresentation is the core issue in political explanations and this is underlined by the *concept of Machiavellianism*. This is the person's tendency to deceive and manipulate others for personal gain (Byrne & Whiten, 1989; Christie & Geis, 1970). The concept is often used to explain cost overruns as a result of competition among parties for government funding or to get projects going. Strategic behaviour is enabled because 'uncertainties of estimates are never brought to the attention of decision-makers' (Odeck, 2004). Similarly, cost overruns can be considered the result of the decision-making process involving many actors with different interests acting strategically (possibly involving 'lying') leading to sub-optimal results. One theory that also incorporates the notion of manipulation is *ethical theory,* which studies the behaviour of people and groups and includes their values, customs and responsibility (Wachs, 1982; LaFolette, 2000). Costs are underestimated because of a lack of loyalty or responsibility to the agent or to a the lack of values in a forecaster's mind to produce accurate figures. Lastly, *agency theory* is also often used to address the strategic behaviour in political explanations. Agency theory (principal agent theory) assumes that people act unreservedly in their own narrowly defined self-interest with, if necessary, guile and deceit (Noreen, 1999). Agency theory can explain why strategic behaviour is made possible by the concept of asymmetric information. It is also used in the context of possible institutional set-ups between parties to guide the decision-making on projects. The asymmetric information makes it possible for an agent to take strategic advantage of the set-up of the funding process to deliberately under-budget their projects in order to see them realised.

Ethical theory is rather specific and its contribution to a full understanding of cost overruns is considered to be small due to its weak relationship with cost overruns. The contribution of the concept of Machiavellianism is mainly related to the manipulation element but this is also incorporated in agency theory by assuming agents act, if necessary, with deceit. Agency theory is therefore held to be the most comprehensive theory. It is considered promising in bringing about a more general understanding of the phenomenon of cost overruns because it can also underlie economic explanations. The relationship between the agent and the principal is characterised by the utility maximising behaviour of agents, hence, the link with the economic causes of cost overruns.

# Conclusions and recommendations

This paper provides an overview of the different explanations for cost overruns; the most commonly used explanations are: economic rational behaviour, strategic behaviour, optimism bias, structure of the organisation, relationship between actors and actors' values and their relationship to the environment. The explanations can be grouped into four different categories: technical explanations, economic explanations, psychological explanations, and political explanations. In addition, the theoretical embeddedness of the explanations was investigated. The extent of the use and the variety of theories used in the literature is actually quite large. Table 4 indicates which theories are considered most appropriate to support the explanations for cost overruns for each category of explanations.

Table 4. Appropriate theories for explaining cost overruns

| *Sub-category of explanations* | *Appropriate theories* |
|---|---|
| ▪ Political explanations | ▪ Machiavellianism |
|  | ▪ Agency theory |
| ▪ Technical explanations | ▪ Forecasting theory |
|  | ▪ Planning theory |
| ▪ Economic explanations | ▪ Neoclassical economics |
|  | ▪ Rational choice theory |
| ▪ Psychological explanations | ▪ Prospect theory |
|  | ▪ Rational choice theory |

Considering the wide variety of explanations and theories, we recommend focusing on the type of explanation before applying a specific theory to better understand the cost overruns in projects. Each type of explanation requires the use of a different theory to understand the way in which cost overruns appeared. Political explanations are the most dominant and agency theory (principal-agent theory specifically) is therefore recommended as a basic theory to understand cost overruns. Agency theory is considered to be the most interesting for the following reasons. First, it is rather specific, and can address cost overruns specifically. Secondly, an initial attempt to use the theory to understand cost overruns has already been made indicating its relevance. And lastly, the theory makes use of several disciplines, including politics, economics and sociology, which makes the theory fairly complete. However, although agency theory is quite comprehensive, it is to be expected that there may be aspects that cannot be addressed appropriately by agency theory. It might not be the all-embracing theory that can be applied to understand and explain cost overruns by political theories. If that is true, an eclectic theory needs to be defined that is based on agency theory but also includes the 'best' insights of other theories. Therefore, the recommendation is to search for other promising theories that can help bring about a better understanding of cost overruns. Theories in the fields of political science, economics or institutions are considered useful. In addition, research into the explanations of cost underestimation with respect to contingencies and explanations regarding demand forecasts is considered valuable.

## Acknowledgement

The research was supported by the Dutch Ministry of Transport. The authors thank two anonymous referees for their useful comments.

## References

Altshuler, A. and D. Luberoff. (2003). *Mega-Projects: The Changing Politics of Urban Public Investment*. Brookings Institution, Washington DC.

Armstrong, J. S. (ed.). (2001). *Principles of forecasting: a handbook for researchers and practitioners*. Kluwer Academic Publishers: Norwell, Massachusetts.

Arrow, K.J. (1987). *Economic theory and the hypothesis of rationality.* The New Palgrave: a dictionary of economics, 2, pp. 69-75.

Arvan, L. and A. P. N. Leite (1990). "Cost overruns in long term projects." *International Journal of Industrial Organisation, 8,* pp.: 443-467.

Auditor General of Sweden, 1994 Riksrevisionsverket. Infrastrukurinvestingar: En Kostnadsjamorelse Mellan Poan of Utyfall i 15 Storre Prosjekt InnomVagverket of Banverket RVV 23

Bruzelius, N., B. Flyvbjerg, W. Rothengatter. (2002). Big decision, big risks. Improving accountability in mega projects. *Transport Policy*, Vol. 9, no. 2, pp. 143–154.

Buehler, R., D. Griffin and H. MacDonald. (1997). The role of motivated reasoning in optimistic time predictions. *Personality and Social Psychology Bulletin,* Vol. 23, no. 3, pp. 238-247.

Cantarelli, C.C., B. Flyvbjerg, B. Wee van, and E.J.E. Molin. Lock-in and its influence on the project performance of large-scale transportation infrastructure projects. Investigating the way in which lock-in can emerge and affect cost overruns. Paper presented at the 88[th] annual meeting of the *Transportation Research Board*, Washington, D.C.

Centraal Bureau voor de Statistiek (CBS) (2005). *Nationale Rekeningen 2005.* CBS, Voorburg/Heerlen

Coleman, J.S. (1992). *Rational choice theory: advocacy and critique.* Sage Publications, Newbury Park.

De Bruijn, H. and M. Leijten. (2007). Megaprojects and contested information. *Transportation Planning and Technology*, Vol. 30, pp 49-69.

Dunleavy, P. (1991). *Democracy, Bureaucracy and Public Choice.* Hemel Hemstead

Faludi, A. (1973). *Planning theory.* Pergamon Press, New York.

Flyvbjerg, B. (1998). *Rationality and power: Democracy in practice*. University of Chicago Press, Chicago.

Flyvbjerg, B., M.K. Skamris Holm and S.L. Buhl. (2002). Underestimating cost in public works. Error or Lie? *Journal of the American Planning Association*, Vol. 68.

Flyvbjerg, B., N. Bruzelius and W. Rothengatter. (2003a). *Megaprojects and Risk: An Anatomy of Ambition*. Cambridge University Press, Cambridge.

Flyvbjerg, B., M.K. Skamris Holm and S.L. Buhl. (2003b). "How Common and How Large Are Cost Overruns in Transport Infrastructure Projects?" *Transport Reviews*, Vol. 23, pp. 71-88.

Flyvbjerg, B. and Cowi, (2004). Procedures for Dealing with Optimism Bias. *Transport Planning: Guidance Document.* UK Department for Transport: London.

Flyvbjerg, B., M.K. Skamris Holm and S.L. Buhl. (2004a). "What Causes Cost Overrun in Transport Infrastructure Projects?". *Transport Reviews*, Vol. 24, no. 1, pp. 3-18.

Flyvbjerg, B. (2005a). "Measuring Inaccuracy in Travel Demand Forecasting: Methodological Considerations Regarding Ramp Up and Sampling." *Transportation Research A*, Vol. 39, no. 6, pp 522-530

Flyvbjerg, B. (2005b). Policy and planning for large infrastructure projects: problems, causes, cures. *World Bank Policy Research Working Paper, WPS3781,* World Bank, Washington DC

Fouracre, P. R., R. J. Allport, and J.M. Thomson. (1990). The performance and impact of rail mass transit in developing countries. Crowthorne, Berkshire, Transport and Road Research Laboratory.

Hall, P. (1980). *Great Planning Disasters.* Penguin Books, Harmondsworth.

Kahneman D. and D. Tversky. (1979). Prospect theory: An analysis of decision under risk. *Econometrica,* Vol. 47, no. 2, pp. 263-292.

Kahneman, D. and C. Lovallo. (1993). Timid choices and bold forecasts: A cognitive perspective on risk taking. *Management Science*, Vol. 39, pp. 17-31.

Kaliba, C., M. Muya, and K. Mumba. (2008). "Cost escalation and schedule delays in road construction projects in Zambia." *International Journal of Project Management*: doi:10.1016/j.ijproman.2008.07.003

Kennisinstituut voor Mobiliteitsbeleid (KIM). (2007). *Mobiliteitsbalans 2007*. KIM: Den Haag

Lee, L. (2008). "Cost Overrun and Cause in Korean Social Overhead Capital Projects: Roads, Rails, Airports, and Ports." Journal of Urban Planning and Development **134**(2): 59-62.

Lovallo, D. and D. Kahneman. (2003). Delusions of success: How optimism undermines executives' decision. *Harvard Business Review*, Vol. 81, pp. 56-63.

Mackie, P. and J. Preston (1998). "Twenty-one sources of error and bias in transport project appraisal." *Transport Policy, 5,* pp 1-7.

Mansfield, N. R., O. O. Ugwu, and T. Doranl. (1994). "Causes of delay and cost overruns in Nigerian construction projects." *International Journal of Project Management, 12, 4,* pp 254-260.

Morris, S. (1990). Cost and Time Overruns in Public Sector Projects. *Economic and Political Weekly,* Vol. 15, pp. 154-168.

Morris, P. and G. H. Hough (1987). *The Anatomy of Major Projects: A Study of the Reality of Project Management.* New York: John Wiley and Sons.

Nijkamp, P., and B. Ubbels. (1999). "How Reliable are Estimates of Infrastructure Costs? A Comparative Analysis," *International Journal of Transport Economics,* Vol. 26, no. 1, pp. 23-53.

NAP. *Completing the Big dig; managing the final stages of Boston's Central Artery/Tunnel Project.* Washington, DC: The National Academies Press; 2003

Noreen, E. (1999). The economics of ethics: a new perspective on agency theory. *Accounting organisations and society,* Vol. 13, pp. 359-369.

Odeck, J. (2004). Cost overruns in road construction? *Transport Policy*, Vol. 11, no. 1, pp. 43–53.

Pickrell, D. (1992) "A Desire Named Streetcar: Fantasy and Fact in Rail Transit Planning," *Journal of the American Planning Association*, Vol. 58, no. 2, pp. 158-176.

Reichelt, K. and J. Lyneis (1999). "The dynamics of project performance: Benchmarking the driers of cost and schedule overrun." *European Management Journal, 17, 2,* pp 135-150.

Schexnayder C., C. Fiori, and S. Weber. Project cost estimating; a synthesis of highway practice, Arizona; 2003.

Snidal, D. (1979). Public goods, Property Rights, and Political organizations. *International Studies Quarterly*, Vol. 23, no. 4, pp. 532-566.

Trujillo, L., E. Quinet, and A. Estache. (2002), Dealing with demand forecasting games in transport privatization. *Transport Policy*, Vol. 9, no. 4, pp. 325-334.

21Van Wee, B. (2007). Large infrastructure projects: a review of the quality of demand forecasts and cost estimations. *Environment and Planning B: Planning and Design*, Vol. 34, pp. 611-625.

Wachs, M. (1982). Ethical Dilemmas in Forecasting for Public Policy. *Public Administration Review,* Vol. 42, pp. 562-557.

Wachs, M. (1987). Forecasts in urban transportation planning: uses, methods and dilemmas. *Climatic Change,* Vol. 11, pp. 61-80.

Wachs, M. (1989). When Planners Lie with Numbers. *Journal of the American Planning Association,* Vol. 55, no. 4, pp. 476-479.